\documentclass[12pt,english,dvipsnames]{article}
\usepackage{ae,aecompl}

\usepackage[T1]{fontenc}
\usepackage[utf8]{inputenc}
\usepackage{geometry}
\usepackage{color}
\usepackage{babel}
\usepackage{units}
\usepackage{mathrsfs}
\usepackage{amsmath}
\usepackage{amssymb}
\usepackage{esint}
\usepackage[unicode=true,pdfusetitle,
 bookmarks=true,bookmarksnumbered=false,bookmarksopen=false,
 breaklinks=false,pdfborder={0 0 1},backref=false,colorlinks=true]
 {hyperref}
\hypersetup{
 linkcolor=Blue,citecolor=Plum, urlcolor=Plum}

\makeatletter
\numberwithin{equation}{section}

\usepackage{babel}
\usepackage{aecompl}
\usepackage{units}
\usepackage{mathrsfs}
\usepackage{breakurl}
\usepackage{xcolor}
\newcommand{\ie}{{\it i.e.,} }
\newcommand{\dr}{{\rm d}}

\makeatother

\begin{document}
\begin{flushright}
FIAN/TD/2015-16\\

\par\end{flushright}

\vspace{0.5cm}

\begin{center}
\textbf{\large{}{Perturbative analysis in higher-spin theories}}
\par\end{center}{\large \par}

\begin{center}
\vspace{1cm}

\par\end{center}

\begin{center}
\textbf{V.E.~Didenko,}\textsc{$^{1}$}\textbf{ N.G.~Misuna}\textsc{$^{2}$}\textbf{
and M.A.~Vasiliev}\textsc{$^{1}$}\\
 \vspace{0.5cm}
 \textsc{$^{1}$} \textit{I.E. Tamm Department of Theoretical Physics,
Lebedev Physical Institute,}\\
 \textit{ Leninsky prospect 53, 119991, Moscow, Russia}\\

\par\end{center}

\begin{center}
\textsc{$^{2}$}\textit{Moscow Institute of Physics and Technology,}\\
 \textit{ Institutsky lane 9, 141700, Dolgoprudny, Moscow region,
Russia}\\

\par\end{center}

\begin{center}
\vspace{0.6cm}
 didenko@lpi.ru, misuna@phystech.edu, vasiliev@lpi.ru \\

\par\end{center}

\vspace{0.4cm}

\begin{abstract}
\noindent A new scheme of the perturbative analysis of the nonlinear
HS equations is developed giving directly the final result for the
successive application of the homotopy integrations which appear in
the standard approach. It drastically simplifies the analysis and
results from the application of the standard spectral sequence approach
to the higher-spin covariant derivatives, allowing us in particular
to reduce multiple homotopy integrals resulting from the successive
application of the homotopy trick to a single integral. Efficiency
of the proposed method is illustrated by various examples. In particular,
it is shown how the Central on-shell theorem of the free theory immediately
results from the nonlinear HS field equations with no intermediate
computations. 
\end{abstract}
\newpage{}

\tableofcontents{}

\section{Introduction}

Conventional $AdS/CFT$ correspondence conjecture \cite{AdSCFT1,AdSCFT2,AdSCFT3}
claims that the bulk action evaluated on equations of motion provides
a generating functional for correlation functions of the boundary
CFT, with boundary values of $AdS$ fields associated with sources
for conformal operators.

General aspects of the $AdS/CFT$ correspondence in higher-spin (HS)
theory were originally discussed in \cite{Sundborg,Witten,Sezginsundell}.
More specifically, Klebanov and Polyakov argued \cite{KlPol} that
nonlinear HS gauge theory is holographically dual to vectorial conformal
theories. The conjecture was further generalized for supersymmetric
case in \cite{LP} and specified for parity preserving HS theories
in \cite{SS}. However, apart from some particular tests at the level
of three-point functions \cite{GY1,GY2} the direct check of this
conjecture is still lacking because an appropriate full nonlinear
action of HS theory is as yet unavailable (see however \cite{BSaction}).
On the other hand some very encouraging results have been obtained
in the one-loop analysis of the HS holography \cite{KlGi,BecTs1,BecTs2}.

Recently an alternative conjecture for the boundary generating functional,
which is essentially different \textit{e.g.} from those of \cite{CS_0form}
(and references therein), has been put forward in \cite{Vasiliev:1504},
where an extension of the original HS system of \cite{more} was proposed
containing a 'Lagrangian form' $\mathcal{L}$ which, upon integration,
gives rise to a functional of modules of solutions of the bulk theory,
\ie of boundary sources, that is gauge invariant under HS gauge transformations.
The extended unfolded equations allow a perturbative reconstruction
of this functional. However, since the construction of \cite{Vasiliev:1504}
involves higher differential forms, the perturbative analysis requires
repeated resolution of higher differential forms in terms of the original
HS fields beyond the linear order. In this case application of the
conventional step-by-step machinery developed in \cite{more} (see
also \cite{Vasiliev:1999ba}) quickly gets complicated leading to
involved expressions hard to operate with.

The aim of this paper is to present a more powerful technique for
perturbative analysis of unfolded HS equations, allowing to evaluate
directly the result of repeated application of homotopy formulae.
Being originally developed for the analysis of the higher-rank forms
of the extended Lagrangian HS system in four dimensions of \cite{Vasiliev:1504}
(more generally, extensions of higher-spin theories to higher differential
forms were considered for instance in \cite{Vasiliev08,BSaction,Frobenius}),
the new approach is useful far beyond this specific problem and, in
particular, for the analysis of higher-order terms in the original
HS theories in various dimensions. As an illustration, we demonstrate
in this paper how the derivation of the standard linearized $4d$
HS equations results from a one-line computation instead of several
pages in the previous approach.

\section{4\textit{d} Nonlinear Higher-Spin Equations\label{sec:HS_eq}}

Known formulations of HS gauge theories in various dimensions \cite{anyD}
have similar structure. For definiteness, in this paper we mainly
consider the historically first example of $4d$ nonlinear HS equations
\cite{more} underlying the $AdS_{4}/CFT_{3}$ HS holography. These
equations have the form 
\begin{eqnarray}
 &  & \mathrm{d}_{X}W+W*W=-i\theta_{\alpha}\wedge\theta^{\alpha}\left(1+\eta B*\varkappa k\right)-i\bar{\theta}_{\dot{\alpha}}\wedge\bar{\theta}^{\dot{\alpha}}\left(1+\bar{\eta}B*\bar{\varkappa}\bar{k}\right),\label{eq:HS_1}\\
\nonumber \\
 &  & \mathrm{d}_{X}B+W*B-B*W=0.\label{eq:HS_2}
\end{eqnarray}
Here $\mathrm{d}_{X}=dx^{\underline{m}}\tfrac{\partial}{\partial x^{\underline{m}}}$
is the space-time de Rham differential (onwards we omit wedge symbol).
$W$ and $B$ are fields of the theory which depend both on space-time
coordinates and on twistor-like variables $Y^{A}=\left(y^{\alpha},\bar{y}^{\dot{\alpha}}\right)$
and $Z^{A}=\left(z^{\alpha},\bar{z}^{\dot{\alpha}}\right)$ where
the spinor indices $\alpha$ and $\dot{\alpha}$ take two values.
The $Y$ and $Z$ variables provide a realization of HS algebra through
the following noncommutative star product 
\begin{equation}
\left(f*g\right)\left(Z,Y\right)=\int d^{4}Ud^{4}Vf\left(Z+U,Y+U\right)g\left(Z-V,Y+V\right)\mathrm{e}^{iU_{A}V^{A}},\label{eq:star_product}
\end{equation}
where $U_{A}V^{A}=U^{A}V^{B}\epsilon_{AB}=u^{\alpha}v^{\beta}\epsilon_{\alpha\beta}+\bar{u}^{\dot{\alpha}}\bar{v}^{\dot{\beta}}\epsilon_{\dot{\alpha}\dot{\beta}}$
and $\epsilon_{AB}=\left(\begin{array}{cc}
\epsilon_{\alpha\beta} & 0\\
0 & \epsilon_{\dot{\alpha}\dot{\beta}}
\end{array}\right)$ is the $sp\left(4\right)$-invariant antisymmetric form built from
the $sp\left(2\right)$-metrics $\epsilon_{\alpha\beta}$, $\epsilon_{\dot{\alpha}\dot{\beta}}$
(integration measure in \eqref{eq:star_product} is implicitly normalized
in such a way that $f*1=f$). It follows then 
\begin{equation}
\left[Y^{A},Y^{B}\right]_{*}=-\left[Z^{A},Z^{B}\right]_{*}=2i\epsilon^{AB},\qquad\left[Y^{A},Z^{B}\right]_{*}=0.
\end{equation}
Inner Klein operators $\varkappa$ and $\bar{\varkappa}$ are specific
elements of the star-product algebra defined as 
\begin{equation}
\varkappa:=\exp\left(iz_{\alpha}y^{\alpha}\right),\qquad\bar{\varkappa}:=\exp\left(i\bar{z}_{\dot{\alpha}}\bar{y}^{\dot{\alpha}}\right)\,.\label{Klein}
\end{equation}
They have the following characteristic properties 
\begin{eqnarray}
 &  & \varkappa*\varkappa=1,\qquad\varkappa*f\left(z^{\alpha},y^{\alpha}\right)=f\left(-z^{\alpha},-y^{\alpha}\right)*\varkappa,\\
\nonumber \\
 &  & f\left(y,z\right)*\varkappa=f\left(-z,-y\right)e^{iz_{\alpha}y^{\alpha}},\label{eq:f*kappa}
\end{eqnarray}
analogously for $\bar{\varkappa}$.

\textbf{$B$ }is a 0-form, while $W$ is a 1-form in a space-time
differential $dx^{\underline{m}}$ and in an auxiliary differential
$\theta^{A}$ dual to $Z^{A}$. All differentials anticommute

\begin{equation}
\left\{ dx^{\underline{m}},dx^{\underline{n}}\right\} =\left\{ dx^{\underline{m}},\theta^{A}\right\} =\left\{ \theta^{A},\theta^{B}\right\} =0.
\end{equation}

In addition to the inner Klein operators of the star-product algebra
there is also a pair of exterior Klein operators $k$ and $\bar{k}$
which have similar properties 
\begin{equation}
kk=1,\qquad kf\left(z^{\alpha};y^{\alpha};\theta^{\alpha}\right)=f\left(-z^{\alpha};-y^{\alpha};-\theta^{\alpha}\right)k,
\end{equation}
(analogously for $\bar{k}$), but as opposed to the inner Klein operators,
$k$ ($\bar{k}$) anticommute with (anti)holomorhic $\theta$ differentials
and hence do not admit a realization in terms of the star-product
algebra. The master fields $B$ and $W$ depend on the exterior Klein
operators yielding a natural splitting of the field content into physical
(propagating) and topological (nonpropagating) sectors of the theory.
The sector of physical fields is represented by $B$ linear in $k$
or $\bar{k}$, while $W$ instead depends on $k\bar{k}$. For topological
sector this is other way around. Note that the topological sector
can be truncated away, while this is not the case for the physical
one. In what follows we consider the propagating fields.

$\eta$ in \eqref{eq:HS_1} is a free complex parameter of the theory
which can be normalized to be unimodular $|\eta|=1$ hence representing
the phase factor freedom. HS theory is parity-invariant in the two
cases of $\eta=1$ ($A$-model) and $\eta=i$ ($B$-model) \cite{SS}.

The generalization of \eqref{eq:HS_1}, \eqref{eq:HS_2} proposed
in \cite{Vasiliev:1504} involves differential forms of higher degrees
and has the form

\begin{eqnarray}
 &  & \!\!\!\!\!\!\!\mathrm{d}_{X}\mathcal{W}+\mathcal{W}*\mathcal{W}=-i\theta_{A}\theta^{A}-i\eta\theta_{\alpha}\theta^{\alpha}\mathcal{B}*\varkappa k-i\bar{\eta}\bar{\theta}_{\dot{\alpha}}\bar{\theta}^{\dot{\alpha}}\mathcal{B}*\bar{\varkappa}\bar{k}+g\delta^{4}\left(\theta\right)*\varkappa k*\bar{\varkappa}\bar{k}+\mathcal{L},\label{eq:HSL_1}\\
\nonumber \\
 &  & \mathrm{d}_{X}\mathcal{B}+\mathcal{W}*\mathcal{B}-\mathcal{B}*\mathcal{W}=0.\label{eq:HSL_2}
\end{eqnarray}
Here $\mathcal{W}$ contains odd-graded differential forms, \ie in
addition to the 1-form $W$ there is a 3-form $\Omega$ with analogous
properties, while $\mathcal{B}$ supports even-graded differential
forms being a sum of the 0-form $B$ and a 2-form $\Phi$. $g$ is
an arbitrary coupling constant and $\delta^{4}\left(\theta\right)=\theta^{\alpha}\theta_{\alpha}\bar{\theta}^{\dot{\alpha}}\bar{\theta}_{\dot{\alpha}}$
is the top form in the $Z$-space. $\mathcal{L}(x)$ is independent
of all variables $Y,Z,k,\bar{k}$ and $\theta$. It is a sum of space-time
2-form $\mathcal{L}_{2}$ and space-time 4-form $\mathcal{L}_{4}$.
$\mathcal{L}_{2}$ is argued to support surface charges on the solutions
such as HS black holes, while $\mathcal{L}_{4}$ is conjectured to
represent a generating functional for correlation functions of the
boundary $CFT$ through $AdS/CFT$ correspondence. The latter is referred
to as the on-shell Lagrangian.

Extension \eqref{eq:HSL_1} and \eqref{eq:HSL_2} of original equations
\eqref{eq:HS_1} and \eqref{eq:HS_2} neither affects the dynamical
field content, nor the structure of interactions. Indeed, system \eqref{eq:HSL_1}-\eqref{eq:HSL_2}
contains \eqref{eq:HS_1}-\eqref{eq:HS_2} as the subsector of the
lowest grade, with all higher-forms perturbatively expressed via $B$
and $W$. The appearance of higher-degree forms significantly complicates
the perturbative analysis necessary to evaluate the on-shell Lagrangian
$\mathcal{L}_{4}$. In practice it forces one to repeatedly process
similar equations, in order to resolve auxiliary $\theta$-dependent
functions in terms of physical ($\theta$-independent) space-time
forms. But straightforward computation quickly gets involved. The
goal of this paper is to develop a general machinery that allows one
to handle perturbatively \eqref{eq:HSL_1}, \eqref{eq:HSL_2} in a
more efficient way.

\section{Perturbative Analysis}

To start a perturbative expansion one has to fix some vacuum solution
to \eqref{eq:HSL_1}, \eqref{eq:HSL_2}. Eq.~\eqref{eq:HSL_2} can
be solved by setting the vacuum value of $\mathcal{B}$ to zero 
\begin{equation}
\mathcal{B}_{0}=0.
\end{equation}
The natural vacuum solution for \eqref{eq:HSL_1} in the 1-form sector
is 
\begin{equation}
W_{0}=\phi_{AdS}+Z_{A}\theta^{A},\label{eq:W_0}
\end{equation}
where $\phi_{AdS}$ is the space-time 1-form of $sp\left(4\right)$-connection
describing the $AdS_{4}$ background 
\begin{eqnarray}
 &  & \phi_{AdS}=-\dfrac{i}{4}\Big(\omega^{\alpha\beta}y_{\alpha}y_{\beta}+\bar{\omega}^{\dot{\alpha}\dot{\beta}}\bar{y}_{\dot{\alpha}}\bar{y}_{\dot{\beta}}+2h^{\alpha\dot{\beta}}y_{\alpha}\bar{y}_{\dot{\beta}}\Big),\\
\nonumber \\
 &  & \mathrm{d}_{X}\phi_{AdS}+\phi_{AdS}*\phi_{AdS}=0.\label{eq:dw_ww}
\end{eqnarray}
(We set the cosmological constant to unity.) It is convenient to extract
the Lorentz connection and vierbein as follows 
\begin{eqnarray}
 &  & \phi_{AdS}=-\frac{i}{4}\phi^{AB}Y_{A}Y_{B}=-\frac{i}{4}\left(\omega^{AB}+h^{AB}\right)Y_{A}Y_{B},\label{eq:w_ads}\\
\nonumber \\
 &  & \omega^{AB}Y_{A}Y_{B}:=\omega^{\alpha\beta}y_{\alpha}y_{\beta}+\bar{\omega}^{\dot{\alpha}\dot{\beta}}\bar{y}_{\dot{\alpha}}\bar{y}_{\dot{\beta}},\label{Lor}\\
\nonumber \\
 &  & h^{AB}Y_{A}Y_{B}:=2h^{\alpha\dot{\beta}}y_{\alpha}\bar{y}_{\dot{\beta}}.\label{trans}
\end{eqnarray}

To proceed, one has to determine vacuum values of $\Omega$ and $\mathcal{L}$,
and then perturbative corrections. At all stages of the computation
one encounters only two types of equations to be solved in the adjoint
and twisted adjoint sectors, namely, 
\begin{equation}
\Delta_{ad}f:=\mathrm{d}_{X}f+\left[\phi_{AdS},f\right]_{*}-2i\mathrm{d}_{Z}f=J,\label{eq:adj_eq}
\end{equation}
or 
\begin{equation}
\Delta_{tw}f:=\mathrm{d}_{X}f-\frac{i}{4}\left[\omega^{AB}Y_{A}Y_{B},f\right]_{*}-\frac{i}{4}\left\{ h^{AB}Y_{A}Y_{B},f\right\} _{*}-2i\mathrm{d}_{Z}f=J,\label{eq:tw_eq}
\end{equation}
where $\mathrm{d}_{Z}:=\theta^{A}\tfrac{\partial}{\partial Z^{A}}$.
Indeed, these equations arise from the expansion of \eqref{eq:HSL_1}
and \eqref{eq:HSL_2}, respectively, around the vacuum with $W_{0}$
from \eqref{eq:W_0} and determine $f$ in terms of some source $J$
which is differential form in $dx^{\underline{m}}$ and $\theta^{A}$
expressed order by order in terms of the lower order fields and/or
differential forms of lower degrees. The $Z$-derivative $\mathrm{d}_{Z}$
emerges from star commutator with $Z_{A}\theta^{A}$ in \eqref{eq:W_0}.
The anticommutator in \eqref{eq:tw_eq} results from the linear dependence
of $\mathcal{B}$ on $k$ or $\bar{k}$ flipping a sign of $h^{AB}Y_{A}Y_{B}$

\begin{equation}
\Delta_{ad}\left(fk\right)=\left(\Delta_{tw}f\right)k.\label{eq:ad-tw}
\end{equation}
This corresponds to the so-called twisted adjoint representation as
opposed to the adjoint one in \eqref{eq:adj_eq}. Our aim is to find
an iterative solution to equations \eqref{eq:adj_eq}, \eqref{eq:tw_eq}
in a closed form.

\section{\label{sec:Homotopy-operators}Homotopy operators}

\subsection{Homotopy trick}

We start with recalling some well-known facts on the standard homotopy
trick. Let a differential $\mathrm{d}$ have the standard property
\begin{equation}
\mathrm{d}^{2}=0\,.\label{eq:dd_0}
\end{equation}
In application to HS theory it will be identified with de Rham differential
in $Z$-space.

Let a homotopy operator $\partial$ obey 
\begin{equation}
\partial^{2}=0.\label{eq:dwdw_0}
\end{equation}
Then the operator 
\begin{equation}
A:=\{\mathrm{d}\,,\partial\}\label{eq:A_d_d}
\end{equation}
obeys 
\begin{equation}
[\mathrm{d}\,,A]=0\,,\qquad[\partial\,,A]=0
\end{equation}
as a consequence of \eqref{eq:dd_0}, \eqref{eq:dwdw_0}. For diagonalizable
$A$ the standard Homotopy Lemma states that cohomology of $\mathrm{d}$,
denoted by $H(\mathrm{d})$, is concentrated in the kernel of $A$ 

\begin{equation}
H(\mathrm{d})\subset KerA\,.\label{eq:hom_lemma}
\end{equation}
In this case the projector $\hat{h}$ to $KerA$ 
\begin{equation}
\hat{h}^{2}=\hat{h}
\end{equation}
exists and obeys 
\begin{equation}
[\hat{h}\,,\mathrm{d}]=[\hat{h}\,,\partial]=0\,.
\end{equation}
Also we can introduce the operator $A^{*}$ such that 
\begin{equation}
A^{*}A=AA^{*}=Id-\hat{h}\,.\label{eq:AA*_Id}
\end{equation}
This allows us to define the operator 
\begin{equation}
\mathrm{d}^{*}:=A^{*}\partial=\partial A^{*}
\end{equation}
that obeys 
\begin{equation}
\mathrm{d}^{*}\mathrm{d}+\mathrm{d}\mathrm{d}^{*}=Id-\hat{h}\,.
\end{equation}
This is equivalent to the resolution of identity 
\begin{equation}
\left\{ \mathrm{d}\,,\mathrm{d}^{*}\right\} +\hat{h}=Id\,.\label{eq:res_id_gen}
\end{equation}
This relation provides a general solution to the equation 
\begin{equation}
\mathrm{d}f=J\label{eq:df_J-1}
\end{equation}
with $\mathrm{d}$-closed $J$ being outside of $H\left(\mathrm{d}\right)$,
$\hat{h}J=0$. For such $J$ from \eqref{eq:res_id_gen} we get 
\begin{equation}
J=\mathrm{d}\mathrm{d}^{*}J,
\end{equation}
so \eqref{eq:df_J-1} turns to 
\begin{equation}
\mathrm{d}\left(f-\mathrm{d}^{*}J\right)=0.
\end{equation}
Hence 
\begin{equation}
f=\mathrm{d}^{*}J+\mathrm{d}\epsilon+g,\label{eq:gen_sol}
\end{equation}
where $\mathrm{d}\epsilon$ is an exact part and $g\in H\left(\mathrm{d}\right)$
remains undetermined.

The above properties are true for the exterior differential $\mathrm{d}=\mathrm{d}_{Z}$
in trivial topology. In this case one can set 
\begin{equation}
\partial=Z^{A}\frac{\partial}{\partial\theta^{A}}\,.
\end{equation}
This gives 
\begin{eqnarray}
 &  & A=\theta^{A}\frac{\partial}{\partial\theta^{A}}+Z^{A}\frac{\partial}{\partial Z^{A}},\\
\nonumber \\
 &  & A^{*}f\left(Z;Y;\theta\right)=\intop_{0}^{1}dt\dfrac{1}{t}f\left(tZ;Y;t\theta\right).
\end{eqnarray}
Checking that this indeed obeys \eqref{eq:AA*_Id} one should use
\begin{equation}
t\dfrac{\partial}{\partial t}f\left(tx\right)=x\dfrac{\partial}{\partial x}f\left(tx\right).\label{eq:A_tx}
\end{equation}

$KerA$ consists of $Z$- and $\theta$-independent functions and
thus, according to Poincaré lemma, relation \eqref{eq:hom_lemma}
turns to an exact equality 
\begin{equation}
H(\mathrm{d}_{Z})=KerA\,.
\end{equation}
Correspondingly, $\hat{h}$ here is simply 
\begin{equation}
\hat{h}J\left(Z;Y;\theta\right)=J\left(0;Y;0\right),
\end{equation}
while 
\begin{equation}
\mathrm{d}_{Z}^{*}J\left(Z;Y;\theta\right)=Z^{A}\dfrac{\partial}{\partial\theta^{A}}\intop_{0}^{1}dt\dfrac{1}{t}J\left(tZ;Y;t\theta\right).\label{eq:dz*}
\end{equation}

Our goal is to extend this consideration to the full covariant derivative
containing both $x$-dependent part and $Z$-dependent part by deriving
an appropriate resolution of identity. Generally, a solution to this
problem can be described in terms of a spectral sequence. However,
in the case of HS equations the final result acquires a particularly
concise form allowing to account a number of terms at once.

\subsection{Spectral sequence analysis}

Let us consider a general system of the form 
\begin{equation}
\Delta f:=\mathrm{d}f+\mathcal{D}f=J\,,\label{eq:gen_eq}
\end{equation}
where $\mathrm{d}$ and $\mathcal{D}$ obey the standard spectral
sequence relations 
\begin{equation}
\mathrm{d}^{2}=0,\qquad\left\{ \mathrm{d},\mathcal{D}\right\} =0\,,\qquad\mathcal{D}^{2}=0\,,\label{eq:DD_Dd_0}
\end{equation}
$f$ is an $m$-form and $J$ is an $\left(m+1\right)$-form obeying
the compatibility condition 
\begin{equation}
\Delta J=0\,.\label{DJ}
\end{equation}

System \eqref{eq:gen_eq} is double-graded with respect to two different
differentials $dx^{\underline{m}}$ and $\theta^{A}$ associated with
$\mathcal{D}$ and $\mathrm{d}$, respectively. The aim of this section
is to write down a formal solution to \eqref{eq:gen_eq} using properties
\eqref{eq:res_id_gen} and \eqref{eq:DD_Dd_0} of $\mathrm{d}$.

Let those components of functions that have degree $n$ in $\theta$
be labelled by the subscript $n$: $f_{n}$, $J_{n}$ etc. We will
assume that the cohomology of $\dr$ is concentrated at the lowest
grade in $\theta$ as is the case for de Rham differential. As $\mathcal{D}$
brings one power of $dx^{\underline{m}}$ while $\mathrm{d}$ brings
one power of $\theta$, we can solve \eqref{eq:gen_eq} sequentially
in powers of $dx^{\underline{m}}$: 
\begin{eqnarray}
 &  & \mathrm{d}f_{m}=J_{m+1},\label{eq:df_J}\\
\nonumber \\
 &  & \mathrm{d}f_{m-1}=J_{m}-\mathcal{D}f_{m},\label{eq:Df_df_J}\\
\nonumber \\
 &  & \mathrm{d}f_{m-2}=J_{m-1}-\mathcal{D}f_{m-1},\\
 &  & ...\nonumber \\
 &  & \mathrm{d}f_{0}=J_{1}-\mathcal{D}f_{1},\label{eq:Df_df_J_2}\\
\nonumber \\
 &  & \mathcal{D}g=J_{0}-\mathcal{D}f_{0}.\,,\label{eq:Dg_J_Df}
\end{eqnarray}
where $g$ is a part of $f$ that belongs to $H\left(\dr\right)$.
Using \eqref{eq:gen_sol}, a solution to \eqref{eq:df_J}-\eqref{eq:Df_df_J_2}
can be put into a compact form (omitting $\Delta$-exact terms)

\begin{equation}
\sum_{n=0}^{m}f_{n}=\sum_{n=0}^{m}\left(-\mathrm{d}^{*}\mathcal{D}\right)^{n}\mathrm{d}^{*}J=\Delta^{*}J\,,\qquad\Delta^{*}J:=\mathrm{d}^{*}\dfrac{1}{1+\mathcal{D}\mathrm{d}^{*}}J\,,\label{eq:delta*}
\end{equation}
where instead of the finite sum we sent $n\to\infty$ since terms
with $n>m$ do not contribute anyway.

Now consider \eqref{eq:Dg_J_Df}. As $g\in H\left(\mathrm{d}\right)$,
the same is true for r.h.s., allowing rewrite the equation as $\mathcal{D}g=\hat{h}\left(J_{0}-\mathcal{D}f_{0}\right),$
and hence, taking into account that $\theta$-dependent terms do not
contribute under the action of $\hat{h}$, 
\begin{equation}
\mathcal{D}g=\hat{h}\left(J-\mathcal{D}\Delta^{*}J\right)\,.\label{eq:Dg_2-1}
\end{equation}
Let $\mathcal{\mathfrak{\mathscr{H}}}$ denote the projector to the
'cohomology source'

\begin{equation}
\mathcal{\mathfrak{\mathscr{H}}}:=\hat{h}\left(1-\mathcal{D}\Delta^{*}\right)=\hat{h}\dfrac{1}{1+\mathcal{D}\mathrm{d}^{*}}.\label{eq:H_adj-1}
\end{equation}
Then \eqref{eq:Dg_2-1} takes the form 
\begin{equation}
\mathcal{D}g=\mathcal{\mathfrak{\mathscr{H}}}(J).\label{eq:Dg_HJ}
\end{equation}

To summarize, a general solution to \eqref{eq:gen_eq} is 
\begin{equation}
f=\Delta^{*}J+\Delta\epsilon+g,\label{eq:gen_sol_adj-1}
\end{equation}
where $\Delta^{*}$ is defined in \eqref{eq:delta*}, $\Delta\epsilon$
represents an exact part of solution, and $g$ solves \eqref{eq:Dg_HJ}.
By this analysis the remaining problem is to solve \eqref{eq:Dg_HJ}.
Since the operators $\Delta^{*}$ and $\mathcal{\mathfrak{\mathscr{H}}}$
provide a generalization of $\mathrm{d}^{*}$ and $\hat{h}$ to the
case with $\mathrm{d}$ extended to $\Delta=\mathrm{d}+\mathcal{D}$
it should not be surprising that an analogue of \eqref{eq:res_id_gen}
holds 
\begin{equation}
\left\{ \Delta,\Delta^{*}\right\} +\mathcal{\mathfrak{\mathscr{H}}}=Id,\label{eq:res_id}
\end{equation}
providing another resolution of identity. To check this, one first
makes use of \eqref{eq:res_id_gen} and, after collecting terms, reduces
\eqref{eq:res_id} to

\begin{equation}
\mathrm{d}^{*}\left(\mathrm{d}\dfrac{1}{1+\mathcal{D}\mathrm{d}^{*}}-\dfrac{1}{1+\mathcal{D}\mathrm{d}^{*}}\mathrm{d}-\dfrac{1}{1+\mathcal{D}\mathrm{d}^{*}}\mathcal{D}\right)=0.\label{eq:d(_)_0}
\end{equation}
Using \eqref{eq:DD_Dd_0} and 
\begin{equation}
\left[\mathcal{D}\mathrm{d}^{*},\mathrm{d}\right]=\mathcal{D}\left(1-\hat{h}\right),
\end{equation}
that follows from \eqref{eq:DD_Dd_0} and \eqref{eq:res_id_gen},
one is left in \eqref{eq:d(_)_0} with terms in brackets, all containing
$\hat{h}$ and containing no $\mathrm{d}$. But from \eqref{eq:res_id_gen}
one concludes that, as $\mathrm{d}$ is proportional to $\theta$,
$\mathrm{d}^{*}$ whatever it is, has to be proportional to $\partial/\partial\theta$.
And as $\hat{h}$ puts $\theta$ to zero, all terms in \eqref{eq:d(_)_0}
vanish because of $\mathrm{d}^{*}$ acting after $\hat{h}$, thus
proving \eqref{eq:res_id}. Moreover, the latter property of $\mathrm{d}^{*}$
also shows that $\mathcal{\mathfrak{\mathscr{H}}}=\hat{h}\dfrac{1}{1+\mathcal{D}\mathrm{d}^{*}}$
is a projector 
\begin{equation}
\mathcal{\mathfrak{\mathscr{H}}}^{2}=\mathcal{\mathfrak{\mathscr{H}}},\label{eq:res_id_1}
\end{equation}
and that 
\begin{equation}
\Delta^{*}\mathcal{\mathfrak{\mathscr{H}}}=0.
\end{equation}
Then from \eqref{eq:res_id_gen} it follows that 
\begin{eqnarray}
 &  & \left\{ \Delta,\Delta^{*}\right\} ^{2}=\left\{ \Delta,\Delta^{*}\right\} ,\\
\nonumber \\
 &  & \left[\mbox{\ensuremath{\mathfrak{\mathscr{H}}}},\Delta\right]=\left[\mbox{\ensuremath{\mathfrak{\mathscr{H}}}},\Delta^{*}\right]=0.\label{eq:res_id_2}
\end{eqnarray}

Let us stress that resolution of unity (\ref{eq:res_id}) is essentially
the central formula in this analysis. In particular, using (\ref{DJ}),
it gives rise to equations (\ref{eq:Dg_HJ}), (\ref{eq:gen_sol_adj-1})
via the substitution 
\begin{equation}
J\equiv\big[\left\{ \Delta,\Delta^{*}\right\} +\mathcal{\mathfrak{\mathscr{H}}}\big]J=\Delta\Delta^{*}J+\mathcal{\mathfrak{\mathscr{H}}}J\,.
\end{equation}

The above analysis was general in nature not using specific properties
of ${\mathcal{D}}$. Being useful it simply accounts the manipulations
involved in the process. Now we are in a position to evaluate specific
operators $\Delta^{*}$ and ${\mathscr{H}}$ that appear in the HS
theory.

\section{\label{sec:HS-spectral-sequence}HS spectral sequence}

Now we set $\mathrm{d}=-2i\mathrm{d}_{Z}$ and $\mathcal{D}$ be either
adjoint or twisted adjoint covariant derivative. We start with the
adjoint case.

\subsection{Adjoint case}

The adjoint $AdS$-derivative is 
\begin{equation}
\mathcal{D}_{ad}=\mathrm{d}_{X}+\left[\phi_{AdS},\bullet\right]_{*}=\mathrm{d}_{X}+\phi^{AB}\left(Y_{A}-i\dfrac{\partial}{\partial Z^{A}}\right)\dfrac{\partial}{\partial Y^{B}},
\end{equation}
Since $Z^{A}\dfrac{\partial}{\partial\theta^{A}}$ is nilpotent, only
such terms of $\mathcal{D}_{ad}$ will contribute to \eqref{eq:delta*},
that differentiate $Z^{A}$ in \eqref{eq:dz*}. They are

\begin{equation}
\mathcal{D}_{ad}\sim-i\phi^{AB}\dfrac{\partial^{2}}{\partial Y^{A}\partial Z^{B}}.\label{eq:D_diff}
\end{equation}

This allows us to rewrite \eqref{eq:delta*} as

\begin{equation}
\Delta_{ad}^{*}J=-\dfrac{1}{2i}Z^{A}\dfrac{\partial}{\partial\theta^{A}}\sum_{n=1}^{\infty}\intop_{\left[0,1\right]^{n}}dt_{1}...dt_{n}\prod_{k=1}^{n}\frac{1}{\left(t_{k}\right)^{k}}\left(-\dfrac{1}{2}\phi^{BC}\dfrac{\partial^{2}}{\partial Y^{B}\partial\theta^{C}}\right)^{n-1}J\left(t_{1}\cdot...\cdot t_{n}Z;Y;t_{1}\cdot...\cdot t_{n}\theta\right).\label{eq:delta*_ad}
\end{equation}
Using the relation

\begin{equation}
\intop_{0}^{1}\intop_{0}^{1}dt_{1}dt_{2}t_{1}^{p}t_{2}^{m}f\left(t_{1}t_{2}x\right)=\dfrac{1}{m-p}\intop_{0}^{1}dt\left(t^{p}-t^{m}\right)f\left(tx\right)\,,\qquad m\neq p\label{eq:homot_reduce}
\end{equation}
the repeated integral in \eqref{eq:delta*_ad} can be reduced to the
single one (an accent next to $\prod$ indicates that the $p=k$ term
is skipped):

\begin{eqnarray}
 &  & \intop_{\left[0,1\right]^{n}}dt_{1}...dt_{n}\prod_{k=1}^{n}\frac{1}{\left(t_{k}\right)^{k}}J\left(t_{1}\cdot...\cdot t_{n}Z;Y;t_{1}\cdot...\cdot t_{n}\theta\right)=\intop_{0}^{1}dt\sum_{k=1}^{n}\left(\prod_{p=1}^{n}\phantom{}'\dfrac{1}{k-p}\right)t^{-k}J\left(tZ;Y;t\theta\right)=\nonumber \\
\nonumber \\
 &  & =\intop_{0}^{1}dt\sum_{k=1}^{n}\dfrac{\left(-1\right)^{n-k}t^{-k}}{\left(k-1\right)!\left(n-k\right)!}J\left(tZ;Y;t\theta\right)=\intop_{0}^{1}dt\sum_{k=1}^{n}\left(\begin{array}{c}
n-1\\
k-1
\end{array}\right)\dfrac{\left(-1\right)^{n-k}t^{-k}}{\left(n-1\right)!}J\left(tZ;Y;t\theta\right)=\nonumber \\
\nonumber \\
 &  & =\dfrac{1}{\left(n-1\right)!}\intop_{0}^{1}\dfrac{dt}{t}\left(\dfrac{1}{t}-1\right)^{n-1}J\left(tZ;Y;t\theta\right).\label{eq:combinat}
\end{eqnarray}
Plugging this back into \eqref{eq:delta*_ad} we arrive at the following
compact expression

\begin{equation}
\Delta_{ad}^{*}J=-\dfrac{1}{2i}Z^{A}\dfrac{\partial}{\partial\theta^{A}}\intop_{0}^{1}dt\dfrac{1}{t}\exp\left(-\dfrac{1-t}{2t}\phi^{BC}\dfrac{\partial^{2}}{\partial Y^{B}\partial\theta^{C}}\right)J\left(tZ;Y;t\theta\right)\label{eq:hexp_ad}
\end{equation}
or equivalently 
\begin{equation}
\Delta_{ad}^{*}J=-\dfrac{1}{2i}Z^{A}\dfrac{\partial}{\partial\theta^{A}}\int_{0}^{1}\frac{dt}{t}J\Big(tZ;Y_{B}-\frac{1-t}{2t}\phi_{B}{}^{C}\frac{\partial}{\partial\theta^{C}};t\theta\Big)\,.\label{inte}
\end{equation}
Let us stress that the integration in (\ref{inte}) is free from singularity
at $t=0$ since every differentiation over $\theta$ brings one power
of $t$.

Now consider $\mathcal{\mathfrak{\mathscr{H}}}_{ad}$ \eqref{eq:H_adj-1}.
As $Z$ is eventually set to zero, only \eqref{eq:D_diff} will contribute,
differentiating the overall factor of $Z$ in \eqref{eq:hexp_ad}.
As a result, \eqref{eq:H_adj-1} can be rewritten as 
\begin{equation}
\mathcal{\mathfrak{\mathscr{H}}}_{ad}J\left(Z;Y;\theta\right)=\hat{h}\left(J+\dfrac{1}{2}\phi^{AB}\dfrac{\partial^{2}}{\partial Y^{A}\partial\theta^{B}}\intop_{0}^{1}dt\dfrac{1}{t}\exp\left(-\dfrac{1-t}{2t}\phi^{CD}\dfrac{\partial^{2}}{\partial Y^{C}\partial\theta^{D}}\right)J\left(tZ;Y;t\theta\right)\right).\label{eq:Dg_3}
\end{equation}
The relation 
\begin{equation}
\dfrac{1}{2}\phi^{AB}\dfrac{\partial^{2}}{\partial Y^{A}\partial\theta^{B}}\exp\left(-\dfrac{1}{2}\phi^{CD}\dfrac{\partial^{2}}{\partial Y^{C}\partial\theta^{D}}\right)=\left[\theta^{A}\dfrac{\partial}{\partial\theta^{A}}+Z^{A}\dfrac{\partial}{\partial Z^{A}},\exp\left(-\dfrac{1}{2}\phi^{CD}\dfrac{\partial^{2}}{\partial Y^{C}\partial\theta^{D}}\right)\right]
\end{equation}
along with \eqref{eq:A_tx} allows us to complete the integration
over $t$ giving the final result 
\begin{equation}
\mathcal{\mathfrak{\mathscr{H}}}_{ad}J\left(Z,Y,\theta\right)=\hat{h}\Big(\exp\left(-\dfrac{1}{2}\phi^{AB}\dfrac{\partial^{2}}{\partial Y^{A}\partial\theta^{B}}\right)J\left(Z;Y;\theta\right)\Big)=\hat{h}\Big(J\left(0;Y_{A}-\dfrac{1}{2}\phi_{A}{}^{B}\frac{\partial}{\partial\theta^{B}};\theta\right)\Big).\label{eq:Had}
\end{equation}

The resolution of identity 
\begin{equation}
\left\{ \Delta_{ad},\Delta_{ad}^{*}\right\} +\mathcal{\mathfrak{\mathscr{H}}}_{ad}=1\label{eq:res_id_ad}
\end{equation}
can be checked directly. To this end one notices that, due to \eqref{eq:A_tx},
$1-\mathcal{\mathfrak{\mathscr{H}}}{}_{ad}$ can be rewritten as 
\begin{equation}
\left(1-\mathcal{\mathfrak{\mathscr{H}}}_{ad}\right)J\left(Z;Y;\theta\right)=\left(\theta^{A}\dfrac{\partial}{\partial\theta^{A}}+Z^{A}\dfrac{\partial}{\partial Z^{A}}\right)\intop_{0}^{1}dt\dfrac{1}{t}\exp\left(-\dfrac{1-t}{2t}\phi^{BC}\dfrac{\partial^{2}}{\partial Y^{B}\partial\theta^{C}}\right)J\left(tZ;Y;t\theta\right)
\end{equation}
and simplifies $\left\{ \Delta_{ad},\Delta_{ad}^{*}\right\} $ with
the help of 
\begin{eqnarray}
 &  & \left\{ Z^{A}\dfrac{\partial}{\partial\theta^{A}},\phi^{BC}\dfrac{\partial^{2}}{\partial Y^{B}\partial Z^{C}}\right\} =\phi^{BC}\dfrac{\partial^{2}}{\partial Y^{B}\partial\theta^{C}},\\
\nonumber \\
 &  & \left\{ Z^{A}\dfrac{\partial}{\partial\theta^{A}},\theta^{B}\dfrac{\partial}{\partial Z^{B}}\right\} =\theta^{A}\dfrac{\partial}{\partial\theta^{A}}+Z^{A}\dfrac{\partial}{\partial Z^{A}},
\end{eqnarray}
and flatness condition \eqref{eq:dw_ww}. An alternative check of
\eqref{eq:res_id_ad} can be carried out via integration by parts
in $\left\{ \Delta_{ad},\Delta_{ad}^{*}\right\} $. Clearly, all relations
\eqref{eq:res_id_1}-\eqref{eq:res_id_2} also hold.

The above derivation of \eqref{eq:hexp_ad} and \eqref{eq:Had} was
based on the explicit evaluation of the geometric progression from
\eqref{eq:delta*}. However there is a simpler way to obtain these
formulas, which will be most useful in the twisted adjoint case and
which we explain now.

Let general equation \eqref{eq:gen_eq} for the adjoint case 
\begin{equation}
-2i\theta^{A}\dfrac{\partial}{\partial Z^{A}}f+\mathrm{d}_{X}f+\phi^{AB}\left(Y_{A}-i\dfrac{\partial}{\partial Z^{A}}\right)\dfrac{\partial}{\partial Y^{B}}f=J\label{eq:adj_eq-1}
\end{equation}
be rewritten as 
\begin{equation}
-2i\left(\theta^{A}+\dfrac{1}{2}\phi^{AB}\dfrac{\partial}{\partial Y^{B}}\right)\dfrac{\partial}{\partial Z^{A}}f=J-\mathcal{D}_{ad}^{Y}f,\label{eq:adj_1}
\end{equation}
where $\mathcal{D}_{ad}^{Y}$ denotes the $Z$-independent part of
$\mathcal{D}_{ad}$ 
\begin{equation}
\mathcal{D}_{ad}^{Y}:=\mathrm{d}_{X}+\phi^{AB}Y_{A}\dfrac{\partial}{\partial Y^{B}}.
\end{equation}
Using that 
\begin{equation}
\left(\theta^{A}+\dfrac{1}{2}\phi^{AB}\dfrac{\partial}{\partial Y^{B}}\right)\dfrac{\partial}{\partial Z^{A}}=\exp\left\{ \dfrac{1}{2}\phi^{BC}\frac{\partial^{2}}{\partial Y^{B}\partial\theta^{C}}\right\} \theta^{A}\dfrac{\partial}{\partial Z^{A}}\exp\left\{ -\dfrac{1}{2}\phi^{BC}\frac{\partial^{2}}{\partial Y^{B}\partial\theta^{C}}\right\} \label{eq:adj_transl}
\end{equation}
and, by virtue of \eqref{eq:dw_ww}, 
\begin{equation}
\left[\mathcal{D}_{ad}^{Y},\exp\left\{ -\dfrac{1}{2}\phi^{BC}\frac{\partial^{2}}{\partial Y^{B}\partial\theta^{C}}\right\} \right]=0\,,
\end{equation}
\eqref{eq:adj_1} is transformed to the form 
\begin{equation}
\mathrm{d}_{Z}\widetilde{f}=\widetilde{J}-\mathcal{D}_{ad}^{Y}\widetilde{f}\,,\label{eq:adj_wave}
\end{equation}
where 
\begin{equation}
\widetilde{f}:=\exp\left\{ -\dfrac{1}{2}\phi^{AB}\frac{\partial^{2}}{\partial Y^{A}\partial\theta^{B}}\right\} f.\label{eq:transf_adj}
\end{equation}
Consistency condition for \eqref{eq:adj_wave} is 
\begin{equation}
\mathrm{d}_{Z}\widetilde{J}+\mathcal{D}_{ad}^{Y}\widetilde{J}=0,
\end{equation}
that together with an obvious fact 
\begin{equation}
\left\{ \mathrm{d}_{Z}^{*},\mathcal{D}_{ad}^{Y}\right\} =0,
\end{equation}
allows us to write down a general solution to \eqref{eq:adj_wave}
as 
\begin{equation}
\widetilde{f}=\mathrm{d}_{Z}^{*}\widetilde{J}+\widetilde{g}+\mathrm{d}_{Z}\widetilde{\epsilon}+\mathcal{D}_{ad}^{Y}\widetilde{\epsilon},
\end{equation}
with arbitrary $\widetilde{\epsilon}$ and $\widetilde{g}$ solving
\begin{equation}
\mathcal{D}_{ad}^{Y}\widetilde{g}=\hat{h}\widetilde{J}.\label{eq:wave_coh_adj}
\end{equation}
The inverse transformation to \eqref{eq:transf_adj} gives a general
solution to \eqref{eq:adj_eq-1} 
\begin{equation}
f=\exp\left\{ \dfrac{1}{2}\phi^{AB}\frac{\partial^{2}}{\partial Y^{A}\partial\theta^{B}}\right\} \mathrm{d}_{Z}^{*}\exp\left\{ -\dfrac{1}{2}\phi^{AB}\frac{\partial^{2}}{\partial Y^{A}\partial\theta^{B}}\right\} J+g+\Delta_{ad}\epsilon\label{eq:delta*_ad_1}
\end{equation}
where, as a consequence of \eqref{eq:wave_coh_adj}, $g$ solves 
\begin{eqnarray}
\mathcal{D}_{ad}^{Y}g & = & \mathscr{H}_{ad}J
\end{eqnarray}
with $\mathscr{H}_{ad}$ \eqref{eq:Had}. Plugging \eqref{eq:dz*}
into \eqref{eq:delta*_ad_1} we obtain \eqref{eq:hexp_ad}.

Finally, resolution of identity \eqref{eq:res_id_ad} follows from
\begin{equation}
\left\{ d_{Z},d_{Z}^{*}\right\} \widetilde{f}+\hat{h}\widetilde{f}=\widetilde{f}\,.
\end{equation}

It may be convenient to use the integral form of formulae \eqref{eq:hexp_ad},
\eqref{eq:Had}. This is achieved through shifting arguments via $\delta$-functions
for bosonic variables (hereafter denoted by Latin letters; as in the
case of star product, the factors of $2\pi$ are included into the
integration measure $dUdV$) 
\begin{equation}
f\left(X\right)=\int dUdV\exp\left\{ iU_{A}\left(V^{A}-X^{A}\right)\right\} f\left(V\right)
\end{equation}
and fermionic variables (hereafter denoted by Greek letters) 
\begin{equation}
f\left(\xi\right)=\int d\zeta d\mu\exp\left\{ \mu_{A}\left(\zeta^{A}-\xi^{A}\right)\right\} f\left(\zeta\right).
\end{equation}
We use Berezin integral conventions with anticommuting symbols $\int d\mu$,
$\int d\zeta$ in order to have commuting unity $\int d\mu\mu=1$
and $\left[\int d\mu\mu,\int d\zeta\zeta\right]=0.$ Then 
\begin{eqnarray}
\triangle_{ad}^{*}J\left(Z;Y;\theta\right) & = & -\dfrac{1}{2i}\int d\mu d\varphi d\chi dUdV\intop_{0}^{1}\dfrac{dt}{t}\exp\left\{ \chi_{A}\varphi^{A}+iU_{A}V^{A}\right\} \cdot\nonumber \\
 &  & \cdot\exp\left\{ \mu t\chi_{A}Z^{A}+\dfrac{i}{2}\left(1-t\right)\phi^{BC}\chi_{B}U_{C}\right\} J\left(tZ;Y+V;t\theta^{A}+\varphi\right),\\
\nonumber \\
\mathfrak{\mathscr{H}}_{ad}J\left(Z;Y;\theta\right) & = & \int d\varphi d\chi dUdV\exp\left\{ \chi_{A}\varphi^{A}+iU_{B}V^{B}+\dfrac{i}{2}\phi^{BC}\chi_{B}U_{C}\right\} J\left(0;Y+V;\varphi\right).
\end{eqnarray}

Let us emphasize that the derivation of these formulae involves evaluation
of multiple homotopy integrals. In the usual approach not using this
machinery, one has to perform analogous manipulations in every specific
perturbative computation, which may be quite hard since 'the wood
may not be seen for the trees'.

\subsection{Twisted adjoint case}

The twisted-adjoint $AdS$-derivative 
\begin{equation}
\mathcal{D}_{tw}=\mathrm{d}_{X}-\dfrac{i}{4}\left[\omega^{AB}Y_{A}Y_{B},\bullet\right]_{*}-\dfrac{i}{4}\left\{ h^{AB}Y_{A}Y_{B},\bullet\right\} _{*}\label{eq:D_tw}
\end{equation}
has the form

\begin{equation}
\mathcal{D}_{tw}=\mathrm{d}_{X}+\omega^{AB}\left(Y_{A}-i\dfrac{\partial}{\partial Z^{A}}\right)\dfrac{\partial}{\partial Y^{B}}-\dfrac{i}{2}h^{AB}\left(Y_{A}Y_{B}-2iY_{A}\dfrac{\partial}{\partial Z^{B}}-\dfrac{\partial^{2}}{\partial Y^{A}\partial Y^{B}}-\dfrac{\partial^{2}}{\partial Z^{A}\partial Z^{B}}\right)\,.\label{eq:D_tw_2}
\end{equation}
In this case a direct evaluation of the geometric progression of \eqref{eq:delta*}
becomes more involved because, instead of \eqref{eq:D_diff}, the
$\nicefrac{\partial}{\partial Z}$-dependent part of $\mathcal{D}_{tw}$
\begin{equation}
\mathcal{D}_{tw}\sim-i\omega^{AB}\dfrac{\partial^{2}}{\partial Y^{A}\partial Z^{B}}-h^{AB}Y_{A}\dfrac{\partial}{\partial Z^{B}}+\dfrac{i}{2}h^{AB}\dfrac{\partial^{2}}{\partial Z^{A}\partial Z^{B}}\label{eq:D_tw_3}
\end{equation}
contains degree $-2$ term in $Z$ destroying homogeneity in the homotopy
parameters $t_{k}$, which for the adjoint case led to the compact
coefficient $\prod_{k=1}^{n}\left(t_{k}\right)^{-k}$ in \eqref{eq:delta*_ad}.
However analysis via the similarity transformation remains applicable.

Once again we rewrite the initial equation 
\begin{equation}
-2i\mathrm{d}_{Z}f+\mathcal{D}_{tw}f=J\label{eq:tw_eq-1}
\end{equation}
as 
\begin{equation}
-2i\left(\theta^{A}+\dfrac{1}{2}\omega^{AB}\dfrac{\partial}{\partial Y^{B}}-\dfrac{i}{2}h^{AB}Y_{B}-\frac{1}{4}h^{AB}\dfrac{\partial}{\partial Z^{B}}\right)\dfrac{\partial}{\partial Z^{A}}f=J-\mathcal{D}_{tw}^{Y}f,\label{eq:tw_1}
\end{equation}
where 
\begin{equation}
\mathcal{D}_{tw}^{Y}:=\mathrm{d}_{X}+\omega^{AB}Y_{A}\dfrac{\partial}{\partial Y^{B}}-\dfrac{i}{2}h^{AB}\left(Y_{A}Y_{B}-\dfrac{\partial^{2}}{\partial Y^{A}\partial Y^{B}}\right).
\end{equation}
Eq. \eqref{eq:tw_1} suggests a proper analogue of Eq.~\eqref{eq:transf_adj}
\begin{equation}
\widetilde{f}:=\exp\left\{ -\dfrac{1}{2}\omega^{AB}\frac{\partial^{2}}{\partial Y^{A}\partial\theta^{B}}+\dfrac{i}{2}h^{AB}Y_{A}\frac{\partial}{\partial\theta^{B}}+\frac{1}{4}h^{AB}\frac{\partial^{2}}{\partial Z^{A}\partial\theta^{B}}\right\} f\label{eq:tw_transf}
\end{equation}
since 
\begin{eqnarray}
 &  & \left(\theta^{A}+\dfrac{1}{2}\omega^{AB}\dfrac{\partial}{\partial Y^{B}}-\dfrac{i}{2}h^{AB}Y_{B}-\dfrac{1}{4}h^{AB}\dfrac{\partial}{\partial Z^{B}}\right)\dfrac{\partial}{\partial Z^{A}}=\nonumber \\
 &  & =\exp\left\{ \dfrac{1}{2}\omega^{AB}\frac{\partial^{2}}{\partial Y^{A}\partial\theta^{B}}-\dfrac{i}{2}h^{AB}Y_{A}\frac{\partial}{\partial\theta^{B}}-\frac{1}{4}h^{AB}\frac{\partial^{2}}{\partial Z^{A}\partial\theta^{B}}\right\} \theta^{C}\dfrac{\partial}{\partial Z^{C}}\cdot\nonumber \\
 &  & \cdot\exp\left\{ -\dfrac{1}{2}\omega^{AB}\frac{\partial^{2}}{\partial Y^{A}\partial\theta^{B}}+\dfrac{i}{2}h^{AB}Y_{A}\frac{\partial}{\partial\theta^{B}}+\frac{1}{4}h^{AB}\frac{\partial^{2}}{\partial Z^{A}\partial\theta^{B}}\right\} .
\end{eqnarray}
This reduces \eqref{eq:tw_eq-1} to 
\begin{equation}
d_{Z}\widetilde{f}=\widetilde{J}-\mathcal{D}_{tw}^{Y}\widetilde{f}
\end{equation}
because, analogously to the adjoint case, 
\begin{equation}
\left[\mathcal{D}_{tw}^{Y},-\dfrac{1}{2}\omega^{AB}\frac{\partial^{2}}{\partial Y^{A}\partial\theta^{B}}+\dfrac{i}{2}h^{AB}Y_{A}\frac{\partial}{\partial\theta^{B}}+\frac{1}{4}h^{AB}\frac{\partial^{2}}{\partial Z^{A}\partial\theta^{B}}\right]=0.
\end{equation}
This allows one to immediately write down a general solution of \eqref{eq:tw_eq-1}
in terms of $f$

\begin{equation}
f=\Delta_{tw}^{*}J+g+\triangle_{tw}\epsilon,
\end{equation}
where

\begin{eqnarray}
 &  & \Delta_{tw}^{*}J=-\dfrac{1}{2i}Z^{C}\dfrac{\partial}{\partial\theta^{C}}\exp\left\{ \dfrac{1}{2}\omega^{AB}\dfrac{\partial^{2}}{\partial Y^{A}\partial\theta^{B}}-\dfrac{i}{2}h^{AB}Y_{A}\frac{\partial}{\partial\theta^{B}}-\dfrac{1}{4}h^{AB}\dfrac{\partial^{2}}{\partial Z^{A}\partial\theta^{B}}\right\} \cdot\nonumber \\
 &  & \cdot\intop_{0}^{1}dt\dfrac{1}{t}\exp\left\{ -\dfrac{1}{2t}\omega^{AB}\dfrac{\partial^{2}}{\partial Y^{A}\partial\theta^{B}}+\dfrac{i}{2t}h^{AB}Y_{A}\frac{\partial}{\partial\theta^{B}}+\dfrac{1}{4t^{2}}h^{AB}\dfrac{\partial^{2}}{\partial Z^{A}\partial\theta^{B}}\right\} J\left(tZ;Y;t\theta\right)\label{eq:hexp_tw_entangl}
\end{eqnarray}
Using Baker-Campbell-Hausdorff formula, we find

\begin{eqnarray}
 &  & \exp\left\{ \dfrac{1}{2}\omega^{AB}\frac{\partial^{2}}{\partial Y^{A}\partial\theta^{B}}-\dfrac{i}{2}h^{AB}Y_{A}\frac{\partial}{\partial\theta^{B}}\right\} =\nonumber \\
 &  & =\exp\left\{ \frac{i}{8}\omega^{AB}\frac{\partial}{\partial\theta^{B}}h_{A}\phantom{}^{C}\frac{\partial}{\partial\theta^{C}}-\dfrac{i}{2}h^{AB}Y_{A}\frac{\partial}{\partial\theta^{B}}\right\} \exp\left\{ \dfrac{1}{2}\omega^{AB}\frac{\partial^{2}}{\partial Y^{A}\partial\theta^{B}}\right\} ,
\end{eqnarray}
that permits to rewrite \eqref{eq:hexp_tw_entangl} as

\begin{eqnarray}
\Delta_{tw}^{*}J & = & -\dfrac{1}{2i}Z^{C}\dfrac{\partial}{\partial\theta^{C}}\intop_{0}^{1}dt\dfrac{1}{t}\exp\left\{ -\dfrac{i}{8}\left(\dfrac{1-t}{t}\right)^{2}\omega^{AB}h_{A}{}^{C}\dfrac{\partial^{2}}{\partial\theta^{B}\partial\theta^{C}}+i\dfrac{1-t}{2t}h^{AB}Y_{A}\dfrac{\partial}{\partial\theta^{B}}\right\} \cdot\nonumber \\
 &  & \cdot\exp\left\{ -\dfrac{1-t}{2t}\omega^{AB}\dfrac{\partial^{2}}{\partial Y^{A}\partial\theta^{B}}+\dfrac{1-t^{2}}{4t^{2}}h^{AB}\dfrac{\partial^{2}}{\partial Z^{A}\partial\theta^{B}}\right\} J\left(tZ;Y;t\theta\right)\label{eq:hexp_tw}
\end{eqnarray}
and $g$ solves 
\begin{equation}
\mathcal{D}_{tw}^{Y}g=\mathscr{H}_{tw}J,
\end{equation}
where 
\begin{eqnarray}
\mathfrak{\mathscr{H}}_{tw}J & := & \hat{h}\exp\left\{ -\dfrac{i}{8}\omega^{AB}h_{A}{}^{C}\dfrac{\partial^{2}}{\partial\theta^{B}\partial\theta^{C}}+\dfrac{i}{2}h^{AB}Y_{A}\dfrac{\partial}{\partial\theta^{B}}\right\} \cdot\nonumber \\
 &  & \cdot\exp\left\{ -\dfrac{1}{2}\omega^{AB}\dfrac{\partial^{2}}{\partial Y^{A}\partial\theta^{B}}+\dfrac{1}{4}h^{AB}\dfrac{\partial^{2}}{\partial Z^{A}\partial\theta^{B}}\right\} J\left(Z;Y;\theta\right).\label{eq:Htw}
\end{eqnarray}
In the same manner as in the adoint case one can check that

\begin{equation}
\left\{ \Delta_{tw},\Delta_{tw}^{*}\right\} +\mathcal{\mathfrak{\mathscr{H}}}_{tw}=1.\label{eq:res_id_tw}
\end{equation}

All general formulae \eqref{eq:res_id_1}-\eqref{eq:res_id_2} remain
valid for $\Delta_{tw}^{*}$ and $\mathcal{\mathfrak{\mathscr{H}}}_{tw}$.
In applications it is useful to use the following integral formulae
(all powers of $2\pi$ are hidden in the integration measure)

\begin{eqnarray}
\triangle_{tw}^{*}J\left(Z;Y;\theta\right) & = & -\dfrac{1}{2i}\int d\mu d\sigma d\rho dUdVdPdQ\intop_{0}^{1}\dfrac{dt}{t}\exp\left\{ \mu t\sigma_{A}Z^{A}+\rho_{A}\sigma^{A}+iU_{A}V^{A}+iP_{A}Q^{A}\right\} \cdot\nonumber \\
 &  & \cdot\exp\left\{ -\dfrac{i}{2}\left(1-t\right)\omega^{AB}\sigma_{A}V_{B}+\dfrac{i}{2}\left(1-t\right)h^{AB}\sigma_{A}\left(Y_{B}+\dfrac{1}{2}U_{B}+\dfrac{1}{2}\left(1+t\right)Q_{B}\right)\right\} \cdot\nonumber \\
 &  & \cdot J\left(tZ+P;Y+U;t\theta+\rho\right),\label{tw1}\\
\nonumber \\
\mathfrak{\mathscr{H}}_{tw}J\left(Z;Y;\theta\right) & = & \int d\sigma d\rho dUdVdPdQ\exp\left\{ \rho_{A}\sigma^{A}+iU_{A}V^{A}+iP_{A}Q^{A}\right\} \cdot\nonumber \\
 &  & \cdot\exp\left\{ -\dfrac{i}{2}\omega^{AB}\sigma_{A}V_{B}+\dfrac{i}{2}h^{AB}\sigma_{A}\left(Y_{B}+\dfrac{1}{2}U_{B}+\dfrac{1}{2}Q_{B}\right)\right\} J\left(P;Y+U;\rho\right).\label{tw2}
\end{eqnarray}
Another convenient representation for \eqref{tw1} and \eqref{tw2}
in the form of integrals over purely fermionic variables{\small{}{}
\begin{align}
 & \Delta_{tw}^{*}J(Z,Y;\theta)=-\frac{1}{2i}Z^{A}\frac{\partial}{\partial\theta^{A}}\int d\nu d\mu\int_{0}^{1}\frac{dt}{t}e^{\nu^{A}\mu_{A}}\cdot\nonumber \\
 & \cdot J\Big(tZ_{A}-\frac{1}{4}(1-t^{2})\nu^{B}h_{AB},Y_{A}+\dfrac{1}{2}(1-t)\nu^{B}\omega_{AB};\mu_{A}+t\theta_{A}-\dfrac{i}{2}(1-t)Y^{B}h_{AB}-\frac{i}{8}(1-t)^{2}\nu^{B}h_{A}{}^{C}\omega_{BC}\Big),\\
 & \mathfrak{\mathscr{H}}_{tw}J\left(Z;Y;\theta\right)=\int d\nu d\mu\,e^{\nu^{A}\mu_{A}}J\left(-\frac{1}{4}\nu^{B}h_{AB},Y_{A}+\dfrac{1}{2}\nu^{B}\omega_{AB};\mu_{A}+\dfrac{i}{2}h_{A}{}^{B}Y_{B}-\frac{i}{8}\nu^{B}h_{A}{}^{C}\omega_{CB}\right)\,
\end{align}
} can be useful in applications.

\section{Applications}

Now let us consider some applications of the developed machinery illustrating
in particular how it immediately reproduces some results which otherwise
would require multi-step computations.

\subsection{Free equations}

Consider a linear analysis of non-extended HS system \eqref{eq:HS_1},
\eqref{eq:HS_2}. In the lowest order it takes the form 
\begin{eqnarray}
\Delta_{ad}W_{1} & = & -i\eta B_{1}*\gamma-i\bar{\eta}B_{1}*\bar{\gamma},\label{eq:lin_1}\\
\nonumber \\
\Delta_{ad}B_{1} & = & 0,\label{eq:lin_2}
\end{eqnarray}
where 1-form $W_{1}$ and 0-form $B_{1}$ are linear fluctuations
and 
\begin{equation}
\gamma:=\theta_{\alpha}\theta^{\alpha}\varkappa k,\qquad\bar{\gamma}:=\bar{\theta}_{\dot{\alpha}}\bar{\theta}^{\dot{\alpha}}\bar{\varkappa}\bar{k}\,.\label{eq:gammas}
\end{equation}
$\gamma$ and $\bar{\gamma}$ are central elements of the HS algebra.
\begin{equation}
\Delta_{ad}\gamma=\left(\Delta_{tw}\varkappa\right)\theta_{\alpha}\theta^{\alpha}k=0,\qquad\Delta_{ad}\bar{\gamma}=\left(\Delta_{tw}\bar{\varkappa}\right)\bar{\theta}_{\dot{\alpha}}\bar{\theta}^{\dot{\alpha}}\bar{k}=0.
\end{equation}
These equations follow from \eqref{eq:ad-tw} as well as from the
fact that Klein operators \eqref{Klein} commute with the Lorentz
part of $AdS$ connection \eqref{Lor} and anticommute with $AdS$
translations \eqref{trans} thus being covariantly constant with respect
to the twisted-adjoint derivative.

As mentioned in Section \ref{sec:HS_eq}, the physical part of $B_{1}$
depends linearly on $k$ or $\bar{k}$ 
\begin{eqnarray}
 &  & B_{1}=Ck+\bar{C}\bar{k},\\
\nonumber \\
 &  & \Delta_{tw}C=\Delta_{tw}\bar{C}=0.
\end{eqnarray}
According to \eqref{eq:gen_sol_adj-1}, $C$ and $\bar{C}$ are $Z$-independent
and obey 
\begin{equation}
\mathcal{D}_{tw}C\left(y,\bar{y}\right)=0\,,\qquad\mathcal{D}_{tw}\bar{C}\left(y,\bar{y}\right)=0.
\end{equation}
Now, in principle, we can find a general solution for $W_{1}$ from
\eqref{eq:lin_1}, using \eqref{eq:gen_sol_adj-1}. But as we are
interested in $\theta$-independent part of the HS equations, where
all dynamical d.o.f. are contained, we can directly look for the $Z$-cohomology
equations with the help of \eqref{eq:Had}. This yields

\begin{equation}
\mathcal{D}_{ad}\omega\left(Y\right)=-i\hat{h}\exp\left(-\dfrac{1}{2}\phi^{AB}\dfrac{\partial^{2}}{\partial Y^{A}\partial\theta^{B}}\right)\left(\eta B_{1}*\gamma+\bar{\eta}B_{1}*\bar{\gamma}\right)
\end{equation}
for $Z$-independent space-time 1-form $\omega$, representing HS
potentials. Elementary computation using \eqref{eq:f*kappa} gives
\begin{equation}
\mathcal{D}_{ad}\omega\left(Y\right)=-\frac{i}{4}\eta h_{\mu}\phantom{}^{\dot{\alpha}}h^{\mu\dot{\beta}}\dfrac{\partial^{2}}{\partial\bar{y}^{\dot{\alpha}}\partial\bar{y}^{\dot{\beta}}}\left(C\left(0,\bar{y}\right)+\bar{C}\left(0,\bar{y}\right)k\bar{k}\right)-\frac{i}{4}\bar{\eta}h^{\alpha}\phantom{}_{\dot{\mu}}h^{\beta\dot{\mu}}\dfrac{\partial^{2}}{\partial y^{\alpha}\partial y^{\beta}}\left(C\left(y,0\right)k\bar{k}+\bar{C}\left(y,0\right)\right).
\end{equation}
So, using the results of Section \ref{sec:HS-spectral-sequence},
we immediately arrive at Central On-Mass-Shell Theorem. Note that
with the new technics one even does not need to know an explicit form
of $W_{1}$.

\subsection{Higher orders}

Let us sketch how the general scheme applies at higher orders.

To the second order, equations \eqref{eq:HS_1}, \eqref{eq:HS_2}
reduce to 
\begin{align}
 & \Delta_{ad}W_{2}+W_{1}*W_{1}=-i\eta B_{2}*\gamma-i\bar{\eta}B_{2}*\bar{\gamma}\,,\\
\nonumber \\
 & \Delta_{tw}B_{2}+\left[W_{1},B_{1}\right]_{*}=0\,,
\end{align}
where $B_{2}$ and $W_{2}$ are master fields at the second order.
The equations for $Z$-independent part (cohomology) of $B_{2}$-field,
$C(Y)k$ and $\bar{C}(Y)\bar{k}$ can be immediately obtained using
\eqref{eq:Dg_HJ} 
\begin{equation}
\mathcal{D}_{tw}C=-\mathcal{\mathfrak{\mathscr{H}}}_{tw}\left(W_{1}*C-C*\pi\left(W_{1}\right)\right),
\end{equation}
where $\pi(y,\bar{y})=(-y,\bar{y})$ and $W_{1}$ is found from \eqref{eq:lin_1}
\begin{equation}
W_{1}=\omega\left(Y\right)-i\Delta_{ad}^{*}\left(\eta B_{1}*\gamma+\bar{\eta}B_{1}*\bar{\gamma}\right).
\end{equation}
The $Z$-dependent part of $B_{2}$ 
\begin{equation}
B_{2}\left(Z;Y\right)=-\Delta_{tw}^{*}\left(W_{1}*B_{1}-B_{1}*W_{1}\right)\label{B2}
\end{equation}
allows one to get the equation for cohomology part of $W_{2}$ 
\begin{equation}
\mathcal{D}_{ad}\omega=-\mathcal{\mathfrak{\mathscr{H}}}_{ad}\left(W_{1}*W_{1}+i\eta B_{2}*\gamma+i\bar{\eta}B_{2}*\bar{\gamma}\right).\label{W2}
\end{equation}
Right hand sides of \eqref{B2} and \eqref{W2} reproduce straightforwardly
second order contribution to HS equations calculated e.g. in \cite{SSan,ST}.

It is not difficult to write down general recurrency for HS equations
at any order in perturbations. Equations for the $n$-th order are
\begin{eqnarray}
 &  & \mathcal{D}_{ad}\omega\left(Y\right)=-\sum_{p=1}^{n-1}\mathcal{\mathfrak{\mathscr{H}}}_{ad}\left(W_{p}*W_{n-p}+i\eta B_{n}*\gamma+i\bar{\eta}B_{n}*\bar{\gamma}\right),\\
\nonumber \\
 &  & \mathcal{D}_{tw}\left(Ck+\bar{C}\bar{k}\right)\left(Y\right)=-\sum_{p=1}^{n-1}\mathcal{\mathfrak{\mathscr{H}}}_{tw}\left(W_{p}*B_{n-p}-B_{n-p}*W_{p}\right),\\
\nonumber \\
 &  & B_{n}\left(Z;Y\right)=-\sum_{p=1}^{n-1}\Delta_{tw}^{*}\left(W_{p}*B_{n-p}-B_{n-p}*W_{p}\right)\,,\\
\nonumber \\
 &  & W_{n}\left(Z;Y\right)=-\sum_{p=1}^{n-1}\Delta_{ad}^{*}\left(W_{p}*W_{n-p}+i\eta B_{n}*\gamma+i\bar{\eta}B_{n}*\bar{\gamma}\right)\,.
\end{eqnarray}
These formulae provide a simplified version of the formulae presented
in \cite{SSan}.

\subsection{Extended equations}

Another application is to extended HS system \eqref{eq:HSL_1}, \eqref{eq:HSL_2}.
As an illustration we calculate a vacuum 3-form $\Omega$, contributing
to $\mathcal{W}$. Corresponding equation, resulting from \eqref{eq:HSL_1},
is 
\begin{equation}
\Delta_{ad}\Omega=g\gamma*\bar{\gamma}.\label{eq:3form_eq}
\end{equation}
First of all we find that the r.h.s. of \eqref{eq:3form_eq} does
not contribute to cohomology 
\begin{equation}
\mathcal{\mathfrak{\mathscr{H}}}_{ad}\left(g\delta^{4}\left(\theta\right)*k\varkappa*\bar{k}\bar{\varkappa}\right)=\hat{h}\Big[\exp\left(-\dfrac{1}{2}\phi^{AB}\dfrac{\partial^{2}}{\partial Y^{A}\partial\theta^{B}}\right)g\exp\left(iZ_{C}Y^{C}\right)\delta^{4}\left(\theta\right)\Big]=0.
\end{equation}
Hence, omitting pure gauge part, we find from \eqref{eq:gen_sol_adj-1}
that 
\begin{align}
 & \Omega=\Delta_{ad}^{*}\left(g\gamma*\bar{\gamma}\right)=-\dfrac{g}{2i}Z^{A}\dfrac{\partial}{\partial\theta^{A}}\intop_{0}^{1}dtt^{3}\exp\left(-\dfrac{i}{2}\left(1-t\right)\phi^{BC}\dfrac{\partial}{\partial\theta^{B}}Z_{C}+itZ_{B}Y^{B}\right)\delta^{4}\left(\theta\right)k\bar{k}=\nonumber \\
 & =-\dfrac{g}{2i}Z^{A}\dfrac{\partial}{\partial\theta^{A}}\intop_{0}^{1}dt\,t^{3}e^{itZ_{A}Y^{A}}\delta^{4}(\theta_{A}-\dfrac{i}{2}(1-t)\phi_{A}{}^{B}Z_{B})k\bar{k}\,,
\end{align}
which coincides with the result presented in \cite{Vasiliev:1504}.
The proposed technique makes its derivation much easier.

For future applications, let us apply resolution of identity \eqref{eq:res_id_tw}
to the central elements $\gamma$ and $\bar{\gamma}$ from \eqref{eq:gammas},
which play an important role in the theory. This yields 
\begin{equation}
\gamma=\Delta_{ad}\rho+\bar{H},\label{eq:gamma_res}
\end{equation}
where 
\begin{eqnarray}
\rho & := & \Delta_{tw}^{*}\gamma=\intop_{0}^{1}dtz^{\alpha}\left(it\theta_{\alpha}+\frac{1}{2}t\left(1-t\right)\omega_{\alpha}{}^{\beta}z_{\beta}-\frac{1}{2}\left(1-t\right)h_{\alpha}{}^{\dot{\beta}}\bar{y}_{\dot{\beta}}\right)\exp\left(itz_{\gamma}y^{\gamma}\right)k,\label{eq:ro}\\
\nonumber \\
\bar{H} & := & \mathcal{\mathfrak{\mathscr{H}}}_{tw}\gamma=\frac{1}{4}h^{\alpha\dot{\beta}}h_{\alpha}{}^{\dot{\gamma}}\bar{y}_{\dot{\beta}}\bar{y}_{\dot{\gamma}}k.
\end{eqnarray}
Analogously 
\begin{eqnarray}
\bar{\gamma} & = & \Delta_{ad}\bar{\rho}+H,\\
\nonumber \\
\bar{\rho} & := & \Delta_{tw}^{*}\bar{\gamma}=\intop_{0}^{1}dt\bar{z}^{\dot{\alpha}}\left(it\bar{\theta}_{\dot{\alpha}}+\frac{1}{2}t\left(1-t\right)\bar{\omega}_{\dot{\alpha}}{}^{\dot{\beta}}z_{\dot{\beta}}-\frac{1}{2}\left(1-t\right)h^{\beta}{}_{\dot{\alpha}}y_{\beta}\right)\exp\left(it\bar{z}_{\dot{\gamma}}\bar{y}^{\dot{\gamma}}\right)\bar{k},\\
\nonumber \\
H & := & \mathcal{\mathfrak{\mathscr{H}}}_{tw}\bar{\gamma}=\frac{1}{4}h^{\beta\dot{\alpha}}h^{\gamma}{}_{\dot{\alpha}}y_{\beta}y_{\gamma}\bar{k}.
\end{eqnarray}

\section{Conclusion}

The tools developed in this paper greatly simplify perturbative analysis
of HS equations. Instead of repeated application of the homotopy formulae
to the equations containing de Rham differential in the twistor variables,
we have derived a closed formula for the application of this procedure
to any function. This approach significantly simplifies the perturbative
computations providing a systematic scheme for the analysis of the
invariant defining system \eqref{eq:HSL_1}, \eqref{eq:HSL_2} which
gets far too annoying in the standard setup where the first-order
perturbation is comparable to the fourth-order of the original system.

The proposed method rests on the application of the general spectral
sequence formula of resolution of identity to specific HS covariant
derivatives. Remarkably, in this case one can write down the explicit
expressions containing a single homotopy integral instead of formal
operator series in terms of homotopy integrals for the 'inverse' operator
and the cohomology projection. Effectively this means that formulae
presented in this paper evaluate most of multiple homotopy integrations
appearing in the conventional computation. This provides an efficient
working tool applicable to a variety of problems far beyond the analysis
of invariant functionals in the $4d$ HS theory \cite{forthc} for
which it has been originally developed.

In particular, the method is applicable to HS models in three \cite{3d}
and any dimensions \cite{anyD}. The application to $d=3$ HS equations
\cite{3d} is straightforward. In this case the structure of adjoint
operator is similar to that in four dimensions \eqref{eq:D_diff}.
The twisted-adjoint operator, though differently realized in $d=3$,
still acts analogously to its $d=4$ cousin \eqref{eq:D_tw}.

In the HS models in any dimensions of \cite{anyD} the situation seems
different as the analog of the four-dimensional $\mathrm{d}_{Z}=\frac{\partial}{\partial Z^{A}}\theta^{A}$
is $\hat{\mathrm{d}}=\frac{\partial}{\partial Z^{I\alpha}}\theta^{I\alpha}$,
where $I$ is a vector index of $o(d-1,2)$ that takes $d+1$ values.
Introduced in a dual fashion to $Y^{I\alpha}$, however, variables
$Z^{I\alpha}$ are almost all excessive. The minimal amount of extra
$Z$-variables is $z_{\alpha}=Z_{d+1\alpha}$. Formally, such reduction
can be easily performed by fixing $\theta$-dependent part of $W$
as $W\sim Z_{i\alpha}\theta^{i\alpha}$ to all orders, where $i\in I$
belongs to Lorentz-transversal part. Upon this 'gauge fixing' the
perturbation series arrange themselves analogously to the $d=4$ form.

\section*{Acknowledgement}

We are grateful to O.Gelfond and E.Skvortsov for a useful comments.
The research was supported by the Russian Science Foundation grant
14-42-00047 in association with Lebedev Physical Institute.


\begin{thebibliography}{10}
\bibitem{AdSCFT1} J. M. Maldacena, \textit{The large N limit of superconformal
field theories and supergravity}, Adv.Theor.Math.Phys. 2 (1998) 231-252
\texttt{\href{http://arxiv.org/abs/hep-th/9711200}{[hep-th/9711200]}}.

\bibitem{AdSCFT2} S. S. Gubser, I. R. Klebanov and A. M. Polyakov,
\textit{Gauge theory correlators from noncritical string theory},
Phys. Lett. B428 (1998) 105-114 \texttt{\href{http://arxiv.org/abs/hep-th/9802109}{[hep-th/9802109]}}.

\bibitem{AdSCFT3} E. Witten, \textit{Anti-de Sitter space and holography},
Adv. Theor. Math. Phys. 2 (1998) 253-291 \texttt{\href{http://arxiv.org/abs/hep-th/9802150}{[hep-th/9802150]}}.

\bibitem{Sundborg} Bo Sundborg, \textit{Stringy gravity, interacting
tensionless strings and massless higher spins}, Nucl.Phys.Proc.Suppl.
102 (2001) 113-119 \texttt{\href{http://arxiv.org/abs/hep-th/0103247}{[hep-th/0103247]}}.

\bibitem{Witten} E. Witten, talk at the John Schwarz 60-th birthday
symposium \texttt{\href{http://theory.caltech.edu/jhs60/witten/1.html}{http://theory.caltech.edu/jhs60/witten/1.html}}.

\bibitem{Sezginsundell} E. Sezgin, P. Sundell, \textit{Massless higher
spins and holography}, Nucl.Phys. B644 (2002) 303-370 \texttt{\href{http://arxiv.org/abs/hep-th/0205131}{[hep-th/0205131}}{]}.

\bibitem{KlPol} I. R. Klebanov and A. M. Polyakov, \textit{AdS dual
of the critical O(N) vector model}, Phys. Lett. B550 (2002) 213-219
\texttt{\href{http://arxiv.org/abs/hep-th/0210114}{[hep-th/0210114]}}.

\bibitem{LP} R. Leigh, A.C. Petkou, \textit{Holography of the N=1
higher spin theory on AdS(4)}, JHEP 0306 (2003) 011, \texttt{\href{http://arxiv.org/abs/hep-th/0304217}{[hep-th/0304217]}}.

\bibitem{SS} E. Sezgin, P. Sundell, \textit{ Holography in 4D (super)
higher spin theories and a test via cubic scalar couplings}, JHEP
0507 (2005) 044 \texttt{\href{http://arxiv.org/abs/hep-th/0305040}{[hep-th/0305040]}}.

\bibitem{GY1} S. Giombi and X. Yin, \textit{Higher Spin Gauge Theory
and Holography: The Three-Point Functions}, JHEP 1009 (2010) 115 \texttt{\href{http://arxiv.org/abs/0912.3462}{[0912.3462]}}.

\bibitem{GY2} S. Giombi and X. Yin, \textit{Higher Spins in AdS and
Twistorial Holography}, JHEP 1104 (2011) 086 \texttt{\href{http://arxiv.org/abs/1004.3736}{[1004.3736]}}.

\bibitem{BSaction} N. Boulanger, P. Sundell, \textit{An action principle
for Vasiliev's four-dimensional higher-spin gravity}, J.Phys. A44
(2011) 495402 \texttt{\href{http://arxiv.org/abs/1102.2219}{[1102.2219]}}.

\bibitem{KlGi} S. Giombi, I. Klebanov, \textit{One Loop Tests of
Higher Spin AdS/CFT}, JHEP 1312 (2013) 068 \texttt{\href{http://arxiv.org/abs/1308.2337}{[1308.2337]}}.

\bibitem{BecTs1} M. Beccaria, A. Tseytlin, \textit{Higher spins in
AdS$_{5}$ at one loop: vacuum energy, boundary conformal anomalies
and AdS/CFT}, JHEP 1411 (2014) 114 \texttt{\href{http://arxiv.org/abs/1410.3273}{[1410.3273]}}.

\bibitem{BecTs2} M. Beccaria, A. Tseytlin, \textit{On higher spin
partition functions}, J.Phys. A48 (2015) 275401 \texttt{\href{http://arxiv.org/abs/1503.08143}{[1503.08143]}}.

\bibitem{CS_0form} N. Colombo, P. Sundell, \textit{Higher Spin Gravity
Amplitudes From Zero-form Charges} \texttt{\href{http://arxiv.org/abs/1208.3880}{[1208.3880]}}.

\bibitem{Vasiliev:1504} M. A. Vasiliev, \textit{Invariant Functionals
in Higher-Spin Theory} \texttt{\href{http://arxiv.org/abs/1504.07289}{[1504.07289]}}.

\bibitem{more} M. A.Vasiliev, \textit{More on equations of motion
for interacting massless fields of all spins in (3+1)-dimensions,
}Phys.Lett. B285 (1992) 225-234.

\bibitem{Vasiliev:1999ba} M. A. Vasiliev, \textit{Higher spin gauge
theories: Star-product and AdS space}, in M.A. Shifman (ed.): The
many faces of the superworld 533-610 \texttt{\href{http://arxiv.org/abs/hep-th/9910096}{[hep-th/9910096]}}.

\bibitem{Vasiliev08} M. A. Vasiliev, \textit{On Conformal, SL(4,R)
and Sp(8,R) Symmetries of 4d Massless Fields}, Nucl.Phys. B793 (2008)
469-526 \texttt{\href{http://arxiv.org/abs/0707.1085}{[0707.1085]}}.

\bibitem{Frobenius} N. Boulanger, E. Sezgin, P. Sundell, \textit{4D
Higher Spin Gravity with Dynamical Two-Form as a Frobenius-Chern-Simons
Gauge Theory} \texttt{\href{http://arxiv.org/abs/1505.04957}{[1505.04957]}}. 

\bibitem{anyD} M. A. Vasiliev, \textit{Nonlinear equations for symmetric
massless higher spin fields in (A)dS(d)}, Phys.Lett. B567 (2003) 139-151
\texttt{\href{http://arxiv.org/abs/hep-th/0304049}{[hep-th/0304049]}}.

\bibitem{SSan} E. Sezgin, P. Sundell, \textit{Analysis of higher
spin field equations in four-dimensions}, JHEP 0207 (2002) 055 \texttt{\href{http://arxiv.org/abs/hep-th/0205132}{[hep-th/0205132]}}.

\bibitem{ST} N. Boulanger, P. Kessel, E.D. Skvortsov, M. Taronna,
\textit{Higher spin interactions in four-dimensions: Vasiliev versus
Fronsdal}, J.Phys. A49 (2016) 095402 \texttt{\href{http://arxiv.org/abs/1508.04139}{[1508.04139]}}.

\bibitem{forthc}V. E. Didenko, N. G. Misuna and M. A. Vasiliev, work
in progress.

\bibitem{3d} S. F. Prokushkin, M. A. Vasiliev, \textit{ Higher spin
gauge interactions for massive matter fields in 3-D AdS space-time},
Nucl.Phys. B545 (1999) 385 \texttt{\href{http://arxiv.org/abs/hep-th/9806236}{[hep-th/9806236]}}.\end{thebibliography}
\end{document}